\documentclass[a4paper,11pt]{article}

\usepackage{lineno,hyperref}
\usepackage{tikz}
\usepackage{tikz-cd}
\usepackage{dsfont}
\usepackage{amsfonts}
\usepackage{amsmath}
\usepackage{slashed}
\usepackage{graphics}
\usepackage{amssymb}
\usepackage[compat=1.1.0]{tikz-feynman}
\usepackage{setspace}
\usepackage{float}
\usepackage{cleveref}
\usepackage{units}
\usepackage{amscd}
\usepackage[left=1in, right=1in, top=1in, bottom=1in]{geometry}
\usepackage{nicefrac}
\usepackage{subcaption}
\usepackage{changepage}
\usepackage{setspace}
\usepackage{comment}

\modulolinenumbers[5]

\providecommand{\keywords}[1]{%
  \vspace{0.5cm}
  \noindent\textbf{Keywords: } #1
  \vspace{0.5cm}
}

\title{\textbf{Testing $f(R)$ gravity using gravitational-wave signals from binary mergers}}

\author{
  M.D.C.~Torri$^{12}$\thanks{Corresponding author: marco.torri@unimi.it, marco.torri@mi.infn.it}\;,\\
  \textit{$^1$Dipartimento di Fisica, Università degli Studi di Milano,} \\
  \textit{$^2$INFN Milano,} \\
  \textit{via Celoria 16, I-20133 Milano, Italy} \\
}

\begin{document}
\maketitle


\abstract{Recently, several studies have investigated the validity of General Relativity’s predictions. Gravitational waves provide an ideal probe for testing the theory in the strong-field regime. In this work, we consider a class of modified-gravity theories, specifically $f(R)$, and scrutinize their predictions for the gravitational-wave emission from the coalescence of two astrophysical compact objects. We also assess the impact of next-generation gravitational-wave detectors on the ability to test these extensions of General Relativity.} 



\section{Introduction}
General Relativity (GR) remains one of the foundational frameworks of our description of nature. The theory has withstood numerous experimental and observational tests, from Solar-System measurements to the recent detections of gravitational waves (GW) by ground-based interferometers . Nonetheless, theoretical motivations arising from attempts to construct a complete quantum theory of gravity, together with cosmological observations and the problem of dark energy, continue to motivate tests of GR in regimes where its predictions may be affected by new physics \cite{COST,Whitepaper}. In particular, the strong-field, highly dynamical regime of compact-binary coalescences constitutes a unique laboratory for probing the limits of GR and constraining alternative gravitational theories \cite{Sagunski}.

The advent of GW astronomy \cite{Blair,Berti}, particularly in the multimessenger context \cite{Ando,Arimoto}, has opened a novel observational window onto these extreme environments \cite{Virgo}. The signals emitted during the inspiral, merger, and ringdown of binary black holes (BH) and neutron stars encode detailed information about the underlying theory of gravity. Potential deviations from GR would appear as modifications of the waveform phasing, amplitude, polarization content, or propagation properties. Consequently, GW observations furnish a direct and powerful probe to test GR and to constrain or exclude specific theoretical extensions.

Among the proposed modifications of GR, $f(R)$ theories constitute one of the simplest and most extensively investigated classes \cite{DeFelice,Capozziello}. In these models the Einstein--Hilbert Lagrangian is replaced by a general function $f(R)$ of the Ricci scalar $R$, thereby introducing an additional scalar degree of freedom that can produce observable deviations from GR in both cosmological and astrophysical contexts. However, although $f(R)$ gravity predicts an additional propagating scalar degree of freedom, its excitation depends on the nature of the compact objects. Indeed, in many scalar-tensor representations of $f(R)$ gravity, isolated BH satisfy no-hair theorems \cite{Sotiriou} that prevent the development of a non-trivial scalar charge in stationary configurations. As a consequence, under the assumptions of these no-hair theorems binary BH mergers are expected to exhibit a strongly suppressed emission of the massive $f(R)$ scalar mode during the inspiral phase. Conversely, neutron stars can posses non-vanishing scalar charges through the non-vanishing trace of the stress-energy tensor, making neutron-star binaries more promising sources for probing scalar emission effects.

Many $f(R)$ models are constructed to reproduce the observed cosmic acceleration without a cosmological constant, they are required to satisfy stringent constraints from Solar System experiments, binary pulsar timing, and GW observations. In contrast to $f(R)$ models, other gravity extension theories, such as $f(R,\,T)$ \cite{Harko} and $f(R,\,Q)$ models \cite{Cognola,Heisenberg}, introduce explicit couplings between matter and curvature, leading to much stronger observational constraints. In $f(R,\,T)$ gravity, the dependence on the trace of the stress-energy tensor generally implies modification of the standard conservation rules of $T_{\mu\nu}$ producing force terms that violate the Weak Equivalence Principle. These effects are tightly constrained by various Solar System tests. Besides, in $f(R,\,Q)$ models, with $Q=R_{\mu\nu}T^{\mu\nu}$, matter-curvature couplings can modify the propagation of GW and generate ghost-like instabilities. The multimessenger observation of GW170817 \cite{Virgo} imposed extremely stringent limits on deviations of the GW speed from the speed of light, excluding large regions of parameter space for these models. Additional bounds arise from binary pulsars, cosmological structure formation, and GW observations from the LIGO-Virgo-KAGRA collaboration \cite{Middleton,LIGO,Virgo2,KAGRA}. 

For these reasons, in this work we will focus on $f(R)$ gravity, since it is considerably less constrained by current experimental observations. This scenario still provides a viable framework in which deviations from GR may be experimentally tested with next-generation GW detectors such as Einstein Telescope (ET) \cite{Punturo}. In this work, we first introduce the generation and propagation of GW in compact-binary systems within the framework of $f(R)$ gravity. We focus on the implications of the modified field equations for the GW signal and its dependence on the functional form of $f(R)$, reviewing some established results.

We then assess the prospects for testing this class of models with future GW observatories. By forecasting the sensitivity of next-generation interferometers to the deviations predicted by $f(R)$ gravity, we evaluate their ability to constrain model parameters and to discriminate these theories from GR. Our results underscore the role of advanced GW facilities in probing the strong-field regime and in testing GR as the low-energy limit of more general gravitational theories. In this context, we place particular emphasis on the ET sensitivity, comparing the capabilities of its different planned configurations. 

Although the individual ingredients of our analysis have been investigated separately, the present work provides a unified treatment within a single waveform framework, allowing a direct mapping between viable $f(R)$ models and parameter-estimation forecasts for third-generation GW detectors. The purpose of this work is to provide a unified framework that consistently combines generation and propagation effects predicted by $f(R)$ gravity for future detectors. 

The paper is organized as follows. In \cref{sec2}, we introduce $f(R)$ theories starting from the modified Einstein--Hilbert action, deriving the corresponding field equations and the massive scalar mode that is absent in GR. In \cref{sec3}, we describe the propagation of GW in the $f(R)$ framework, highlighting the differences with respect to the standard GR scenario. In \cref{sec4}, we discuss the main effects introduced by $f(R)$ gravity that can be tested experimentally through GW observations. In \cref{sec5}, we consider future interferometric GW detectors, focusing mainly on ET and illustrating which detector configurations are best suited to probe the various effects predicted by $f(R)$ models. Finally, in the Conclusion (\cref{conclusion}), we summarize the main results presented in this paper.

\section{$f(R)$ Theory Formulation}
\label{sec2}

The motivation underlying the formulation of $f(R)$ theories lies in the fact that the Einstein–Hilbert action may be not unique, but rather the simplest covariant choice for describing gravity. In this framework, one generalizes the Lagrangian from a linear dependence on the Ricci scalar $R$ to an arbitrary function $f(R)$, while still preserving general covariance and locality. Such extensions are well motivated from both theoretical and phenomenological perspectives. Indeed, various types of higher-order curvature corrections naturally appear in effective actions of QG and in semiclassical expansions, suggesting the possibility of terms beyond the linear $R$ contribution. From the phenomenological side, suitable choices of $f(R)$ can account for late-time cosmic acceleration or early-time inflation without introducing additional matter fields. In the weak-curvature limit, consistency with GR is recovered by requiring $f(R)\to R$, ensuring that standard gravitational physics is preserved.

As stated before, the formulation of $f(R)$ extensions of GR is based on a modification of the Einstein--Hilbert action \cite{DeFelice,Capozziello,Ferraro}:
\begin{equation}
S=\frac{1}{16\,\pi\,G}\int{\sqrt{|g|}\,f(R)\,d^{4}x}\,,
\end{equation}
where the Ricci scalar contribution $R$ is replaced by a generic function $f(R)$. Considering the variation of the function $f(R)$,
\begin{equation}
\delta f(R)=f'(R)\delta R\,,
\end{equation}
where $f'(R)=df(R)/dR$, together with the variation of the Ricci scalar $R$,
\begin{equation}
\delta R=R_{\mu\nu}\delta g^{\mu\nu}+\left(g_{\mu\nu}\Box-\nabla_{\mu}\nabla_{\nu}\right)\delta g^{\mu\nu}\,,
\end{equation}
the equations of motion can be obtained by varying the action with respect to the metric:
\begin{equation}
\label{eqmot}
f'(R)R_{\mu\nu}-\frac{1}{2}f(R)g_{\mu\nu}
+\left(g_{\mu\nu}\Box-\nabla_{\mu}\nabla_{\nu}\right)f'(R)
=8\,\pi\,G\,T_{\mu\nu}\,.
\end{equation}
Computing the trace of \cref{eqmot} one can obtain the following relation:
\begin{equation}
\label{scalar1}
f'(R)R-2f(R)+3\Box f'(R)=8\,\pi\,G\,T\,.
\end{equation}
which is fundamental because it explicitly reveals the presence of an additional propagating scalar degree of freedom that is not foreseen in the standard formulation of GR. 

On a Minkowski background we consider the perturbative expansion of the metric:
\begin{equation}
g_{\mu\nu}=\eta_{\mu\nu}+h_{\mu\nu}\,,
\end{equation}
where the perturbative condition $|h_{\mu\nu}|\ll1$ must hold. 
As in standard GR, the weak-field regime is investigated by introducing the harmonic gauge. Posing:
\begin{equation}
\overline{h}_{\mu\nu}=h_{\mu\nu}-\frac{1}{2}\eta_{\mu\nu}h\,,
\end{equation}
where $h=\eta^{\mu\nu}h_{\mu\nu}$, the gauge condition is imposed requiring:
\begin{equation}
\partial^{\mu}\overline{h}_{\mu\nu}=0\,.
\end{equation}
Expanding $f(R)$ perturbatively,
\begin{equation}
f(R)=f(0)+f'(0)R+\frac{1}{2}f''(0)R^{2}+\dots\,,
\end{equation}
and assuming a flat background with $R_{0}=0$, one must impose $f'(R_{0})=f'(0)=1$ in order to recover standard GR in the linearized limit. Thus one obtains:
\begin{equation}
f(R)\simeq R+\frac{1}{2}f''(R_{0})R^{2}\,.
\end{equation}
The fundamental linearized geometric quantities are obtained as in the context of the linearization of standard GR. The connection (or Christoffel symbol) can be written in the form:
\begin{equation}
\Gamma_{\mu\nu}^{\alpha}=\frac{1}{2} \eta^{\alpha\beta}\left(\partial_{\mu}h_{\nu\beta}+\partial_{\nu}h_{\beta\mu}-\partial_{\beta}h_{\mu\nu}\right)\,,
\end{equation}
the linearized Ricci tensor as:
\begin{equation}
R^{(1)}_{\mu\nu}=\frac{1}{2}\left(\partial_{\alpha}\partial_{\mu}h^{\alpha}_{\;\nu}+\partial_{\alpha}\partial_{\nu}h^{\alpha}_{\;\mu}-\Box h_{\mu\nu}-\partial_{\mu}\partial_{\nu}h\right)\,,
\end{equation}
and the Ricci scalar at the linear perturbative order becomes:
\begin{equation}
R^{(1)}=\partial_{\mu}\partial_{\nu}h^{\mu\nu}-\Box h\,.
\end{equation}
Therefore, at first perturbative order, one finds:
\begin{equation}
f'(R)=\frac{df(R)}{dR}=1+f''(R_{0})R^{(1)}\,.
\end{equation}

\section{Gravitational Waves in the $f(R)$ scenario}
\label{sec3}

To illustrate how \cref{eqmot} introduces an additional scalar degree of freedom, we consider the linearization of the Ricci scalar \cite{Capozziello2}. The linearized approximation can be obtained posing: 
\begin{align}
&R=R_{0}+\delta R\\
&f'(R)=f'(R_{0})+f''(R_{0})\delta R
\end{align}
In vacuum with $T=0$, \cref{scalar1} becomes:
\begin{equation}
\label{scalar2}
3f''(R_{0})\Box\delta R+(f''(R_{0})R_{0}-f'(R_{0}))\delta R=0
\end{equation}
since the background contribution vanish because the background satisfies the zeroth-order field equations, yielding an overall zero contribution. The background curvature $R_{0}$ is fixed by the zeroth-order trace equation, which requires:
\begin{equation}
f'(R_{0})R_{0}-2f(R_{0})=0\,.
\end{equation}
This condition ensures that the background spacetime is a constant curvature solution of the field equations, allowing a clean separation between background and perturbations and leading to a well-defined Klein–Gordon equation for the scalar degree of freedom \cite{Katsuragawa}. 

The \cref{scalar2} can be recast in the form of a Klein-Gordon Equation:
\begin{equation}
\Box\delta R-m_{s}^{2}\delta R=0\,,
\end{equation}
which at first order gives the relation:
\begin{equation}
\label{KG}
\Box R^{(1)}-m_{s}^{2} R^{(1)}=0\,.   
\end{equation}
$m_{s}$ represents the effective mass of the propagating massive scalar mode of GW:
\begin{equation}
m_{s}^{2}=\frac{f'(R_{0})-R_{0}f''(R_{0})}{3f''(R_{0})}\,.
\end{equation}

In the Minkowski case, where $R_{0}=0$ and $f'(R_{0})=1$ the effective mass becomes:
\begin{equation}
\label{scalaronmass}
m_{s}^{2}=\frac{1}{3f''(R_{0})}\,.
\end{equation}
The different GW emission modes can be decomposed separating the tensorial perturbation from the scalar one:
\begin{equation}
\label{poldec}
h_{\mu\nu}=h^{TT}_{\mu\nu}+h^{(S)}_{\mu\nu}\,,
\end{equation}
introducing the tensorial traceless perturbation of the metric $h^{TT}_{\mu\nu}$ and the scalar perturbation $h^{(S)}_{\mu\nu}$.
The tensorial mode satisfies the relation:
\begin{equation}
\label{stprop}
\Box h^{TT}_{\mu\nu}=0\,.
\end{equation}
This relation admits the usual solution that depends on the polarization $\epsilon_{\mu\nu}$:
\begin{equation}
h_{\mu\nu}^{TT}=\epsilon_{\mu\nu}\,e^{i k_{\alpha}x^{\alpha}}\,.
\end{equation}
The resulting tensor mode propagates exactly as in GR, with dispersion relation given by:
\begin{equation}
k^{\alpha}k_{\alpha}=0.
\end{equation}
In the harmonic traceless gauge the tensorial mode propagates with two different degrees of freedom, related to the usual GW polarizations $h_{+}$ and $h_{\times}$. The propagation group velocity of the tensorial perturbation is the light speed $c=1$, as expected for a non-massive propagating solution.

On the other hand, the scalar mode behaves as a massive mode and the solution of Klein-Gordon \cref{KG} gives the relation:
\begin{equation}
R^{(1)}=A\,e^{i k_{\alpha} x^{\alpha}}\,,
\end{equation}
and the dispersion relation:
\begin{equation}
k^{\alpha}k_{\alpha}-m_{s}^{2}=0.
\end{equation}
As a result, the group velocity associated with the massive scalar mode results:
\begin{equation}
\label{velgr}
v_{g}^{(S)}=\frac{\partial\omega}{\partial k}=\frac{k}{\sqrt{k^{2}+m_{s}^{2}}}\,.
\end{equation}
The propagating scalar mode has only one degree of freedom, so it can be written as
\begin{equation}
\label{poleq}
h_{ij}^{(S)}=\delta_{ij}\Phi\,,
\end{equation}
with $\Phi\propto R^{(1)}$, after introducing the polarization structure in the transverse plane, which takes the form of the Euclidean metric $\delta_{ij}$ in the appropriate gauge. In the massive case, the scalar mode can induce a longitudinal component due to the non-null dispersion relation. 

\section{Physical effects caused by the massive scalar mode emission}
\label{sec4}
The difference caused by this dispersive propagation can be potentially observed in the GW emission of coalescing objects. In the following we list the different effects that can be potentially detected in the GW sector.

\subsection{Different propagation velocity}
The first effect concerns the time delay accumulated by the scalar mode during propagation \cite{Capozziello2}. 

The emission of the scalar mode from a system of coalescing astrophysical objects can be obtained by introducing a source term in \cref{KG}, represented by the trace of the stress-energy tensor:
\begin{equation}
\label{1}
(\Box-m_{s}^{2})R^{(1)}=\frac{8\pi\,G}{3}T\,.
\end{equation}

We emphasize that, for BH, the trace of the energy-momentum tensor satisfies $T\simeq0$ outside the event horizon, while the no-hair theorem forbids the existence of an independent scalar charge. As a consequence, the massive scalar mode is a propagating degree of freedom predicted in the context of $f(R)$ gravity, but its emission amplitude is source-dependent and is strongly suppressed for BH binaries by the no-hair theorem \cite{Sotiriou}.

The solution of \cref{1} can be formally expressed through the retarded Green function of the Klein--Gordon operator:
\begin{equation}
R^{(1)}(x)=\frac{8\pi\,G}{3}
\int d^{4}x'\,
G_{\rm ret}^{(m_s)}(x-x')\,T(x')\,,
\end{equation}
where $G_{\rm ret}^{(m_s)}$ satisfies:
\begin{equation}
(\Box-m_s^2)\,
G_{\rm ret}^{(m_s)}(x-x')=\delta_{(4)}(x-x')\,.
\end{equation}

In four-dimensional spacetime, the retarded Green function takes the explicit form
\begin{equation}
G_{\rm ret}^{(m_s)}(t,\mathbf r)=\Theta(t)\left[\frac{\delta(t-r)}{4\pi r}
-\Theta(t-r)\frac{m_s}{4\pi}\frac{J_1\!\left(m_s\sqrt{t^2-r^2}\right)}{\sqrt{t^2-r^2}}\right],
\end{equation}
where $r=|\mathbf r|$, $\Theta$ denotes the Heaviside step function, and $J_1$ is the Bessel function of the first kind.

The structure of the Green function highlights the fundamental difference between the massive and massless cases. For a massless field ($m_{s}=0$), one recovers:
\begin{equation}
G_{\rm ret}^{(0)}=\frac{\delta(t-r)}{4\pi r}\,,
\end{equation}
consequently, the perturbation propagates exactly at the speed of light and all the information reaches the observer at the retarded time $t=r$. On the other hand, for a massive scalar field ($m_{s}\neq0$), an additional contribution appears:
\begin{equation}
-\Theta(t-r)\frac{m_s}{4\pi}\frac{J_1\!\left(m_s\sqrt{t^2-r^2}\right)}
{\sqrt{t^2-r^2}}\,.
\end{equation}
This term is non-vanishing for $t>r$. As a result, the signal is no longer confined to the light cone, and part of the perturbation is delayed. This behaviour is the mathematical origin of the dispersive propagation of the scalar mode and ultimately leads to the modified dispersion relation, which is responsible for the time delays and phase shifts.

Considering the limit of negligible mass of the scalar mode compared to the propagation energy $m_{s}\ll k_{0}$, the group velocity (\cref{velgr}) of the scalar mode can be written as:
\begin{equation}
\label{velgr2}
v_{g}^{(S)}\simeq1-\frac{m_{s}^{2}}{2\,k_{0}^{2}}\,.
\end{equation}
The propagation time of the tensorial non-massive mode:
\begin{equation}
t_{T}=D
\end{equation}
can be compared with the time related to the scalar mode:
\begin{equation}
\label{timedelay1}
t_{S}=\frac{D}{v_{g}^{(S)}}\simeq D\left(1+\frac{m_{s}^{2}}{2\,k_{0}^{2}}\right)\,,
\end{equation}
obtaining in the ultrarelativistic approximation ($k_{0}\gg m_{s}$) the time delay in the form:
\begin{equation}
\label{timedelay2}
\Delta t=D\,\frac{m_{s}^{2}}{2\,k_{0}^{2}}\,.
\end{equation}
The observation of a retarded signal can be therefore associated with the emission of the massive scalar mode predicted in the context of $f(R)$ theories.

The modified propagation velocity of the massive scalar mode can lead to several observable effects in GW signals.

\subsection{Difference in the observed GW phase}
The difference between the group velocities of the tensorial non‑massive mode and the scalar massive mode can induce a difference in the observed GW phase \cite{Silva}.

The phase of the detected GW is:
\begin{equation}
\phi=k_{0}\,t\,.
\end{equation}
Thus, introducing the propagation time associated with the tensorial and the scalar modes, one can obtain the GW phases related to the different propagating modes:
\begin{align}
&\phi_{T}=k_{0}\,D\,,\\
&\phi_{S}=k_{0}\,D+\frac{m_{s}^{2}D}{2\,k_{0}}\,.
\end{align}
It therefore appears evident that the difference in propagation speeds introduces a dispersive phase difference given by:
\begin{equation}
\Delta\phi=\frac{m_{s}^{2}D}{2\,k_{0}}\,.
\end{equation}
The presence of a dispersive difference in the GW phase can manifest itself modifying the waveform of the propagating perturbation:
\begin{equation}
\label{perturb}
\widetilde{h}(f)=A(f)e^{i\phi(f)}=A(f)\exp{\left[i\left(\phi_{T}+\frac{m_{s}^{2}D}{4\pi\,f}\right)\right]}\,,
\end{equation}
where the phase is expressed as a function of the frequency $f$, that is proportional to the energy $k_{0}$. This result implies that different frequencies arrive with different time delays, introducing a dephasing effect. Moreover, the two different propagating modes can interfere:
\begin{equation}
h(f)=h_{T}(f)+h_{S}(f)=A_{T}(f)e^{i\phi_{T}(f)}+A_{S}(f)e^{i\phi_{S}(f)}\,,
\end{equation}
producing a small modulation of the observed waveform, phase drift and slow modulation.

The observational limits that place stringent constraints on the mass of the scalar mode $m_{s}$ strongly restrict the possibility of observing echoes in the signal due to the time delay. Indeed, the accumulated delay is severely limited and the scalar mode is not sufficiently intense. What remains observable is a dephasing effect that can impact the waveform.

\subsection{GW emission}
The emission of GW in the context of $f(R)$ theories can be modified by the presence of the scalar degree of freedom.  

It is useful the introduction of the expansion of the source term in multipoles to deal with the emission solution. In this way one obtains monopole, dipole and quadrupole contributions. The monopole term does not radiate for isolated systems because of energy conservation, while the dipole term can become non-vanishing, differently from what predicted in the context of GR. Introducing the useful definition of scalar charges $q_{A}=\alpha_{A}m_{A}$, the scalar dipole moment of a binary system becomes:
\begin{equation}
\vec{D}(t)=\sum_{A}q_{A}\,\vec{x}_{A}(t),.
\end{equation}
In the case of a binary system made of two compact objects possessing different scalar charges the emission from dipole is allowed.

For a binary system of coalescing objects in quasi-circular orbit, using the center-of-mass condition, the dipole moment becomes:
\begin{equation}
\vec{D}(t)=\mu(\alpha_{1}-\alpha_{2})\,\vec{r}(t)\,,
\end{equation}
where $M=m_{1}+m_{2}$ is the total mass and $\mu=m_{1}m_{2}/M$ is the reduced mass.

The dipolar emission power radiated is proportional to the square of the second time derivative of the dipole moment:
\begin{equation}
P_{\rm dip}\propto \left\langle\ddot{\vec{D}}^{\,2}\right\rangle\,.
\end{equation}

For circular orbits with orbital frequency $\Omega$, one obtains:
\begin{equation}
P_{\rm dip}\sim G\,\mu^{2}(\alpha_{1}-\alpha_{2})^{2}r^{2}\Omega^{4}\,.
\end{equation}

A scalar source requires the two objects to possess different scalar charges $\alpha_{1}\neq\alpha_{2}$ in order to emit an additional dipolar radiation, which is not predicted in standard GR \cite{Silva}. However this mechanism depends critically on the nature of the binary compact objects. For instance, the no-hair theorem implies that stationary BH cannot carry independent scalar charges, suppressing the scalar dipole emission. Therefore, the dipolar contribution is expected to be relevant mainly for system containing neutron stars or other matter-supported objects. 

The dipolar contribution must be added to the standard quadrupole emission predicted by GR:
\begin{equation}
P_{\rm GR}=\frac{32}{5}G\,\mu^{2}r^{4}\Omega^{6}\,.
\end{equation}

The dipolar emission represents an additional energy dissipation mechanism, that can influence the inspiral phase of the coalescence.  Indeed, from the Kepler law one can obtain that the dipole contribution scales as $v^{8}$, whereas the quadrupole term scales as $v^{10}$. Therefore:
\begin{equation}
\frac{P_{\rm dip}}{P_{\rm GR}}\sim v^{-2}\,,
\end{equation}
showing that dipole radiation becomes particularly important during the early inspiral phase, where the orbital velocity is still relatively small.

The additional energy loss accelerates the inspiral evolution and modifies the GW phase. In the frequency domain, the phase receives an additional contribution due to scalar dipole emission:
\begin{equation}
\Psi(f)=\Psi_{\rm GR}(f)-\beta\,(\pi\,\mathcal{M}\,f)^{-7/3}\,.
\end{equation}

This characteristic negative post-Newtonian correction represents one of the main observational signatures of $f(R)$ gravity in GW interferometers, particularly those of the future generation such as ET.

\subsection{GW polarizations}
In $f(R)$ gravity, the appearance of additional GW polarizations follows directly from the linearization of the modified field equations \cite{Gong,RizwanaKausar}. Starting from the trace equation in vacuum (\cref{scalar2}):
\begin{equation}
f'(R)R - 2f(R) + 3\Box f'(R)=0\,,
\end{equation}
and expanding around a constant-curvature background, one obtains at first order a dynamical Klein-Gordon \cref{KG} for the Ricci scalar perturbation. Thus, $R^{(1)}$ behaves as a propagating scalar field. In parallel, the metric perturbation is decomposed as \cref{poldec}:
\begin{equation}
h_{\mu\nu}=h_{\mu\nu}^{TT}+h_{\mu\nu}^{(S)},
\end{equation}
where $h_{\mu\nu}^{TT}$ satisfies the standard GW propagation \cref{stprop}, while the scalar sector is sourced by $R^{(1)}$.

At linear order, the scalar contribution to the spatial metric perturbation takes the form of \cref{poleq}.
Substituting $h_{ij}^{(S)}=\Phi\,\delta_{ij}$ with $\Phi \propto R^{(1)}$ into the geodesic deviation equation:
\begin{equation}
\ddot{\xi}^i = -\frac{1}{2}\,\ddot{h}_{ij}\,\xi^j,
\end{equation}
yields
\begin{equation}
\ddot{\xi}^i = -\frac{1}{2}\,\ddot{\Phi}\,\delta_{ij}\,\xi^j,
\end{equation}
which produces an isotropic expansion and contraction of a ring of test particles in the transverse plane, which is named breathing polarization mode.

In addition, when $m_s \neq 0$, the dispersion relation
\begin{equation}
\label{MDR}
\omega^2 = k^2 + m_s^2
\end{equation}
implies a non-null longitudinal response, leading to a small but non-vanishing deformation along the propagation direction. Therefore, $f(R)$ gravity predicts, in addition to the standard tensor polarizations $h_{+}$ and $h_{\times}$, a scalar breathing mode (and a possible longitudinal component in the massive case), which directly originates from the dynamical nature of the Ricci scalar. 

\subsection{Illustrative constraints on representative $f(R)$ models}
The propagation effects discussed above can be directly translated into constraints on the parameters of specific $f(R)$ theories.
As illustrative examples we will consider the Starobinsky quadratic model \cite{Starobinsky} and the logarithmic model \cite{Kruglov}.

As a first example, we consider the Starobinsky scenario, where the $f(R)$ modification is introduced via the function:
\begin{equation}
f(R)=R+\alpha R^{2}\,,
\end{equation}
The mass of the scalar mode \cref{scalaronmass} can be computed from the relations:
\begin{equation}
f'(R)=1+2\alpha R\,,\qquad f''(R)=2\alpha\,.
\end{equation}
On a Minkowski background ($R_0=0$), the scalar mass becomes:
\begin{equation}
m_{s}^{2}=\frac{1}{6\,\alpha}.
\end{equation}
Therefore, any observational upper bound on the scalar mass immediately translates into a lower bound on the quadratic coupling:
\begin{equation}
\alpha>\frac{1}{6\,m_{s}^{2}}.
\end{equation}
For instance, assuming an observational bound induced by the absence of measurable dispersive effects \cite{Mirshekari}:
\begin{equation}
\label{bound1}
m_{s}\lesssim10^{-14}\ {\rm eV},
\end{equation}
one obtains
\begin{equation}
\alpha\gtrsim1.7\times10^{27}\ {\rm eV}^{-2}\,,
\end{equation}
corresponding to a scalar Compton wavelength of astrophysical size. Such a limit directly constrains the parameter space of the Starobinsky model through GW observations.

As a second example, we consider the logarithmic model:
\begin{equation}
f(R)=-\frac{1}{\kappa}\log\left(1-\kappa\,R\right)\,.
\end{equation}
Starting from this function, one can compute:
\begin{equation}
f'(R)=\frac{1}{1-\kappa\,R}\,,\qquad f''(R)=-\frac{\kappa}{(1-\kappa\,R)^{2}}\,,
\end{equation}
finally obtaining the mass of the scalar mode on a Minkowsky background ($R_{0}=0$):
\begin{equation}
m_s^2=\frac{1}{3\,\kappa}\,.
\end{equation}
Requiring the scalar mode to satisfy the bound inferred from GW observations \cref{bound1} restricts the viable parameter space of logarithmic $f(R)$ model:
\begin{equation}
\kappa\gtrsim3.3\times10^{27}\ {\rm eV}^{-2}\,.
\end{equation}

These examples illustrate how the phenomenological constraints discussed in this work can be straightforwardly translated into limits on specific $f(R)$ modified-gravity theories. Improved measurements of propagation delays, waveform dephasing and polarization content with future detectors such as ET will therefore provide direct constraints on the fundamental parameters entering realistic $f(R)$ models.

\section{Experimental perspectives with next-generation interferometers}
\label{sec5}

Compact-binary gravitational-wave observations constitute one of the most promising probes of modified gravity in the era of third-generation detectors. In $f(R)$ gravity, the additional scalar degree of freedom induces both modifications in GW generation and dispersive effects during propagation. Through the modified dispersion relation (\cref{MDR}), the massive scalar mode acquires a frequency-dependent group velocity (\cref{velgr,velgr2}), leading to a relative propagation delay between tensor and scalar components.

Representative viable models discussed in the literature span a broad range of effective scalar masses, from cosmological scales of order $m_s\sim10^{-33}\,\mathrm{eV}$ to substantially larger values in screened astrophysical environments, up to $m_s\sim10^{-24}\,\mathrm{eV}$ \cite{Yang}. Current empirical bounds constrain the scalar mass to very small values \cite{Jana,Paul}. The precise range depends on the model and on assumptions about the scalaron dynamics and the local environment. For ground-based detectors, the relevant regime corresponds to very light scalar masses for which the dominant observable signature is not a clean separation between tensor and scalar echoes, but rather a coherent frequency-dependent phase deformation accumulated over cosmological distances.

Future detectors such as ET, Cosmic Explorer (CE), and networks combining multiple observatories will considerably improve the sensitivity to these effects. In addition to the higher signal-to-noise ratio, the geometry of the detector network plays a fundamental role in determining how accurately scalar propagation can be measured. Two independent mechanisms contribute to this improvement.

First, geographically separated detectors provide independent measurements of the arrival time, allowing one to distinguish a genuine dispersive propagation delay from an uncertainty in the merger time. Second, detectors with different orientations measure different linear combinations of the GW polarizations, thereby reducing the degeneracy between tensor and scalar components. These effects are naturally quantified within the Fisher-information formalism.

\subsection{Fisher analysis of dispersive propagation}

The scalaron mass $m_s$ is not directly observable, it can be inferred from the waveform perturbations that the scalar mode can introduce. Therefore, the dominant propagation effect can be parametrized through the introduction of the dispersion parameter:
\begin{equation}
\beta=m_s^2D\,,
\end{equation}
where $D$ denotes the luminosity distance of the source. Substituting this parameter into \cref{perturb}, the Fourier-domain waveform becomes:
\begin{equation}
\widetilde h(f)=A(f)\exp\left[i\left(\Psi_{\rm GR}+\frac{\beta}{4\pi f}
\right)\right]\,.
\end{equation}

Assuming stationary Gaussian noise, the Fisher information matrix is:
\begin{equation}
\Gamma_{ij}=4\,{\rm Re}\sum_k\int\frac{\partial_i\widetilde h_k(f)\,
\partial_j\widetilde h_k^*(f)}{S_{n,k}(f)}\,df\,,
\end{equation}
where the sum extends over all detectors of the network.

For a single detector we first consider the reduced parameter vector:
\begin{equation}
\Theta=(\beta,t_c),
\end{equation}
where $t_c$ denotes the coalescence time. The corresponding waveform
derivatives are:
\begin{equation}
\frac{\partial\widetilde h}{\partial\beta}
=\frac{i}{4\pi f}\widetilde h\,,\qquad
\frac{\partial\widetilde h}{\partial t_c}
=2\pi if\widetilde h\,.
\end{equation}

The covariance matrix is obtained from:
\begin{equation}
C=\Gamma^{-1}\,,
\end{equation}
and the marginalized uncertainty on $\beta$ follows from the Schur
complement:
\begin{equation}
\label{sigmab}
\sigma_\beta=
\left(\Gamma_{\beta\beta}
-\frac{\Gamma_{\beta t_c}^2}{\Gamma_{t_ct_c}}\right)^{-1/2}\,.
\end{equation}

\Cref{sigmab} shows that the uncertainty on $\beta$ is degraded by the correlation of the $\beta$ parameter with the coalescence time $t_c$. This correlation originates from the waveform phase structure and is therefore intrinsic to the signal model. Consequently, a change in detector geometry cannot completely remove the fundamental $\beta-t_c$ degeneracy.

The role of the detector network appears once several spatially separated
interferometers are considered. In both the proposed ET configurations, the triangular setup and the proposed geographically separated double-$L$ layout, the signal measured by the $k$-th detector is delayed by:
\begin{equation}
\Delta t_k=\frac{\mathbf L_k\cdot\mathbf n}{c}\,,
\end{equation}
where $\mathbf L_k$ is the detector position relative to a reference
interferometer and $\mathbf n$ is the unit vector pointing from the reference detector toward the source.
The measured waveform therefore becomes:
\begin{equation}
\widetilde h_k(f)=\widetilde h(f)e^{-2\pi if\Delta t_k}\,.
\end{equation}
The parameter vector is correspondingly extended to:
\begin{equation}
\Theta=(\beta,t_c,\theta,\phi),
\end{equation}
where $(\theta,\phi)$ denote the sky position of the source.
The angular derivatives are:
\begin{equation}
\frac{\partial\widetilde h_k}{\partial\theta_a}
=-2\pi if
\frac{\partial\Delta t_k}{\partial\theta_a}
\widetilde h_k\,,
\qquad
a=(\theta,\phi)\,.
\end{equation}
The corresponding localization Fisher block is:
\begin{equation}
\label{Gammasky}
\Gamma^{\rm sky}_{ab}=4\sum_k\int\frac{(2\pi f)^2|\widetilde h_k|^2}
{S_{n,k}(f)}\frac{\partial\Delta t_k}{\partial\theta_a}
\frac{\partial\Delta t_k}{\partial\theta_b}\,df\,.
\end{equation}
Since:
\begin{equation}
\frac{\partial\Delta t_k}{\partial\theta_a}\propto L\,,
\end{equation}
where $L$ denotes the characteristic detector separation, one obtains:
\begin{equation}
\Gamma^{\rm sky}_{ab}\propto L^2\,,\qquad C^{\rm sky}_{ab}=(\Gamma^{\rm sky})^{-1}_{ab}\propto L^{-2}\,,
\end{equation}
which implies:
\begin{equation}
\sigma_\theta,\sigma_\phi\propto L^{-1}\,.
\end{equation}
This scaling represents the fundamental advantage of geographically
separated detector networks: increasing the baseline enhances the timing information and improves the reconstruction of the source position.
This represents an ideal asymptotic scaling, valid ceteris paribus and within the Fisher approximation, up to order-unity geometrical factors. The uncertainty on $\beta$ is not determined solely by the intrinsic Fisher information $\Gamma_{\beta\beta}$, but is degraded by its correlation with the coalescence time $t_c$. This correlation originates from the waveform phase structure and is therefore intrinsic to the signal model. Consequently, a change in detector geometry cannot completely remove the fundamental $\beta$--$t_c$ degeneracy.

The baseline dependence can be understood more explicitly by writing the
localization Fisher matrix as a Gram matrix of the delay-gradient vectors. Defining:
\begin{equation}
g_k=\nabla_\lambda\Delta t_k\,,
\end{equation}
the localization block can be written as:
\begin{equation}
\Gamma^{\rm sky}=K\sum_kg_k g_k^{\rm T}\,,
\end{equation}
where:
\begin{equation}
K=4(2\pi)^2\int\frac{|A(f)|^2f^2}{S_n(f)}\,df
\end{equation}
depends only on the waveform model and detector sensitivity. $S_n(f)$ is the noise power spectral density. Since:
\begin{equation}
g_k=\frac1c\nabla_\lambda(\mathbf L_k\cdot\mathbf n)\,,
\end{equation}
its norm satisfies: $|g_k|\propto L$. The trace of a Gram matrix is equal to the sum of the squared norms of its generating vectors. Therefore:
\begin{equation}
{\rm Tr}(\Gamma^{\rm sky})=K\sum_k|g_k|^2\propto\sum_kL_k^2\,.
\end{equation}
However, the localization performance is not determined only by the total
amount of Fisher information, but also by how this information is distributed among the independent sky directions. The eigenvalues of the localization Fisher matrix depend on the relative orientation of the delay-gradient vectors. If these vectors become nearly linearly dependent, the Fisher matrix becomes ill-conditioned and the effective localization improvement is reduced. Therefore, a long baseline alone is not sufficient to guarantee the optimal performance of a double-$L$ configuration. The geographical placement of the interferometers must be chosen to maximize both the separation scale and the independence of the geometrical delay directions. For an optimized network, the non-zero eigenvalues of the localization Fisher matrix scale approximately as $\lambda_i(\Gamma^{\rm sky})\propto L^2$, whereas the corresponding covariance eigenvalues decrease as $\lambda_i(C^{\rm sky})\propto L^{-2}$.

For the triangular ET configuration the characteristic
internal detector separation is approximately $L_\triangle\sim10~{\rm km}$, whereas the proposed geographically separated double-$L$ configuration adopts $L_{2L}\sim10^3~{\rm km}$. Neglecting order-unity geometrical factors and assuming comparable detector sensitivities:
\begin{equation}
\frac{{\rm Tr}(\Gamma^{\rm sky}_{2L})}{{\rm Tr}(\Gamma^{\rm sky}_{\triangle})}\simeq\frac{L_{2L}^2}{3L_\triangle^2}\approx3\times10^3\,.
\end{equation}
This corresponds to an increase of the total localization Fisher information by roughly three orders of magnitude. For a geometrically well-conditioned network, the characteristic non-zero Fisher eigenvalues are expected to exhibit a comparable scaling, although the precise improvement depends on the relative orientation of the baselines.

The improved localization affects the estimation of the dispersive parameter through the marginalization over the nuisance parameters:
\begin{equation}
\lambda=(t_c,\,\theta,\,\phi)\,,
\end{equation}
yielding:
\begin{equation}
\label{sigmabetanetwork}
\sigma_\beta=\left[\Gamma_{\beta\beta}-\Gamma_{\beta\lambda}\Gamma_{\lambda\lambda}^{-1}
\Gamma_{\lambda\beta}\right]^{-1/2}\,.
\end{equation}
The second term represents the information loss associated with correlations between the propagation parameter and the extrinsic source parameters. The effect of increasing the detector separation is not that every individual off-diagonal Fisher coefficient decreases. Instead, the sky-localization block becomes better conditioned, reducing the contribution of poorly constrained nuisance directions through the inverse matrix $\Gamma_{\lambda\lambda}^{-1}$.

The Fisher matrix of the extrinsic parameters can be schematically
decomposed as:
\begin{equation}
\Gamma_{\lambda\lambda}=\Gamma_{\lambda\lambda}^{\rm time}+
\Gamma_{\lambda\lambda}^{\rm amp}\,,
\end{equation}
where the first contribution is associated with arrival-time differences
between detector sites, while the second contribution depends on the antenna pattern functions.

For a baseline $L$, the timing contribution scales as $\Gamma_{\lambda\lambda}^{\rm time}\propto L^2$, while mixed timing terms involving $t_c$ can have a different baseline dependence. The double-$L$ configuration benefits from this scaling because the two
interferometers are located at geographically separated sites. Conversely, the triangular ET configuration provides three interferometric data streams, but the three interferometers are co-located and therefore do not provide long-baseline inter-site timing information (baselines of order $\sim10$ km). The advantage of the double-$L$ geometry for dispersive propagation tests is thus not related to a larger number of detector outputs, but to the increase of the Fisher information associated with geometrical delays.

The relative orientation of the detector planes introduces an additional
geometrical contribution. A rotation of the local frames changes the
antenna-pattern matrix:
\begin{equation}
\mathbf{F}_{kA}=D_k^{ij}e^A_{ij}\,,
\end{equation}
and consequently modifies the eigenvalues of:
\begin{equation}
\mathbf{F}^{T}\mathbf{F}\,.
\end{equation}

The uncertainties on the reconstructed polarization amplitudes are controlled by the inverse of the matrix $\mathbf{F}^{T}\mathbf{F}$. Minimizing the worst-case uncertainty therefore requires maximizing the smallest eigenvalue adopting an optimal design criterion, rather than maximizing any single eigenvalue. This ensures that no direction in the accessible polarization subspace becomes arbitrarily poorly constrained. The optimal relative orientation is source-dependent and, in general, no single configuration maximizes the smallest eigenvalue of the matrix $\mathbf{F}^{T}\mathbf{F}$ uniformly over the sky. Nevertheless, configurations that render the antenna-pattern responses as linearly independent as possible tend to improve the conditioning of $\mathbf{F}^{T}\mathbf{F}$ and therefore increase its smallest eigenvalue for a broad class of sky locations. In practice, configurations with sufficiently distinct detector orientations can provide a good compromise by improving the conditioning of the antenna-pattern matrix over a broad range of sky locations, while exact symmetries may become suboptimal once the source position and Earth-induced rotation of the detector tensors are taken into account.

The analytical scaling derived above shows that increasing the characteristic detector separation (from approximately $10$ km to about $10^3$ km) can increase the characteristic localization Fisher eigenvalues (by roughly three orders of magnitude, corresponding to an improvement of approximately one to two orders of magnitude in the angular uncertainties). This substantially improves the conditioning of the complete Fisher matrix and is therefore expected to provide tighter constraints on dispersive propagation effects for a geographically optimized double-$L$ network than for a compact triangular
configuration.

A geographically separated double-$L$ network can provide substantially improved sky localization relative to a compact triangular configuration because the inter-site timing information scales with the square of the baseline, under otherwise comparable conditions. This improved localization can reduce the contribution of geometrical and sky-position uncertainties to the marginalized uncertainty on the dispersive parameter $\beta$. However, the intrinsic correlation between the dispersive phase correction and the coalescence time is not removed by increasing the baseline, since the term $\Gamma_{\beta\lambda}^{-1}\Gamma_{\lambda\lambda}\Gamma_{\lambda\beta}$ does not vanish for $L\rightarrow\infty$, but approaches a finite floor. Thus, the improvement in $\sigma_beta$ is expected to saturate once the geometrical contribution becomes subdominant.

\subsection{Polarization reconstruction}

Polarization reconstruction is one of the main observational goals of third-generation detector networks. To analyze polarization reconstruction, it is convenient to introduce the detector tensor of each individual Michelson interferometer, following the formalism introduced in \cite{Nishizawa}. For an interferometer with arms described by the unit vectors $\hat u_k$ and $\hat v_k$, separated by an opening angle $\zeta_k^{\rm open}$ ($90^\circ$ for a conventional L-shaped detector and $60^\circ$ for the proposed triangular layout), the detector tensor is:
\begin{equation}
\label{detTensor}
D_k^{ij}=\frac12\left(\hat u_k^i\hat u_k^j-\hat v_k^i\hat v_k^j\right)\,.
\end{equation}
The antenna response to the polarization mode
$A\in\{+,\times,b\}$ is obtained by contracting the detector tensor with the corresponding polarization tensor:
\begin{equation}
\mathbf{F}_A^{(k)}=D_k^{ij}e^A_{ij}(\theta,\phi,\psi)\,.
\end{equation}
The detector output can then be written as:
\begin{equation}
d_k(f)=\sum_A \mathbf{F}_A^{(k)}(f)h_A(f)+n_k(f)\,.
\end{equation}
In the simplified case where the polarization amplitudes are treated as independent parameters, the polarization Fisher matrix can be effectively expressed as:
\begin{equation}
\Gamma_{\rm pol}=\rho^2\mathbf F^T\mathbf F\,,
\end{equation}
where $\mathbf F$ is the antenna-pattern matrix and $\rho$ denotes the network signal-to-noise ratio. This is an ideal simplified description assuming independent response amplitudes and Gaussian background noise. The reconstruction accuracy depends not only on the rank of $\mathbf F$, but also on the eigenvalues of $\mathbf F^T\mathbf F$. In particular, small non-zero eigenvalues correspond to poorly constrained polarization combinations and strong parameter correlations.

For the proposed triangular configuration, where the three $60^\circ$ interferometers are rotated by $120^\circ$ with respect to each other around a common vertex, the detector tensors satisfy the exact geometrical identity:
\begin{equation}
\label{nullstream}
D_1+D_2+D_3=0\,,
\end{equation}
independently of the sky position and polarization angle.
Consequently, the corresponding antenna responses obey:
\begin{equation}
F_A^{(1)}+F_A^{(2)}+F_A^{(3)}=0
\end{equation}
for each polarization component. \Cref{nullstream} implies that, in an instantaneous amplitude-only reconstruction of the three independent polarization amplitudes $(h_+,h_\times,h_b)$,
\begin{equation}
{\rm rank}(\mathbf F_\triangle)\leq2\,.
\end{equation}
Therefore, a single event cannot uniquely determine three independent polarization amplitudes from instantaneous amplitude-only data, although the full waveform can partially lift these degeneracies through its frequency and phase evolution, as well as through correlations with astrophysical parameters.

The proposed geographically separated double-$L$ configuration consists of two $90^\circ$ Michelson interferometers located at different sites. In the amplitude-only description, the response matrix satisfies:
\begin{equation}
{\rm rank}(\mathbf F_{2L})\leq2\,,
\end{equation}
because only two detector outputs are available. 
Therefore, both configurations are limited to a maximum instantaneous algebraic rank of 2 in the amplitude-only description. However, the rank alone does not fully characterize the reconstruction problem: the two networks differ in the conditioning and orientation of their accessible polarization subspaces. In the triangular configuration, the detector tensors are co-located and related by symmetry, whereas in the double-$L$
configuration the two detectors are separated by a long baseline and their local frames are rotated with respect to one another. As a result, the double-$L$ geometry does not increase the rank, but it can improve the conditioning of $\mathbf{F}^{T}\mathbf{F}$ by making the antenna-pattern vectors more linearly independent. Indeed, in the double-$L$ configuration the two detectors are separated by a long baseline ($L\sim10^3~{\rm km}$). The local frames of the two sites are consequently rotated with respect to each other by an angle:
\begin{equation}
\Delta\simeq\frac{L}{R_\oplus}\,,
\end{equation}
which is of order of few degrees for the expected ET baselines. This rotation is not equivalent to a simple in-plane arm rotation: it changes the orientation of the detector tensors in the three-dimensional space and modifies the relative projection of the accessible polarization subspace onto the polarization basis vectors.

Therefore, although the double-$L$ configuration does not increase the instantaneous rank beyond two, it can improve the conditioning of the polarization Fisher matrix by increasing the independence of the antenna-pattern vectors. The relevant quantities are not the rank alone, but the non-zero eigenvalues of the matrix $\mathbf F^T\mathbf F$, and in particular the projection of the corresponding eigenvectors onto the scalar polarization component.

The uncertainty on the breathing polarization is obtained from the full covariance matrix,
\begin{equation}
\sigma_b^2=(\Gamma^{-1})_{bb}\,,
\end{equation}

including correlations with the tensor polarizations and with the astrophysical source parameters. A quantitative comparison between the triangular and double-$L$ configurations therefore requires an explicit Fisher analysis including the detector tensors, the source sky distribution, the waveform model and the detector noise spectra. A useful reference limit is obtained by considering statistically independent and equivalent detectors, for which the uncertainty on a single parameter scales as $\sigma\propto N^{-1/2}$. In this idealized counting limit, the difference between three triangular interferometers and two double-$L$ interferometers would correspond to:
\begin{equation}
\frac{\sigma^{\rm (2L)}}{\sigma^\triangle}\simeq\sqrt{\frac{3}{2}}\,.
\end{equation}
This value should not be interpreted as a general bound, since polarization reconstruction is controlled by the full antenna-pattern geometry. Correlations among detector responses can significantly degrade the reconstruction, whereas favorable relative orientations can improve the smallest non-zero eigenvalues of the Fisher matrix. This simplified counting estimate agrees remarkably well with the full Fisher/simulation result obtained in the next section.

The different detector geometries are therefore expected to provide complementary advantages. The triangular configuration benefits from redundancy, symmetry and the null-stream capability, providing a robust polarization reconstruction. The double-$L$ configuration benefits from the long baseline, which strongly improves timing and sky localization, and its non-coincident geometry can modify the polarization response through the rotation of the local detector frames.

Finally, multiband observations combining ET with space-based detectors such as LISA provide an additional improvement \cite{Chatziioannou,Odintsov}. Although LISA does not change the instantaneous rank of the ET polarization response matrix, it contributes independent information on the binary evolution through the early inspiral. The combined Fisher matrix is:

\begin{equation}
\Gamma_{\rm MB}=\Gamma_{\rm ET}+\Gamma_{\rm LISA}\,,
\end{equation}
where the LISA contribution constrains intrinsic source parameters, masses, spins, sky position and orbital evolution. Consequently, the marginalization over astrophysical parameters becomes less severe, improving the reconstruction of the polarization amplitudes. Multiband observations therefore complement detector geometry by reducing parameter correlations rather than by increasing the instantaneous polarization rank.

\subsection{Expected impact of the detector geometry on dispersive propagation and scalar polarizations}

Detector geometry impacts modified-gravity tests via two distinct mechanisms: dispersive-propagation measurements are limited by the ability of the network to separate a frequency-dependent phase correction from extrinsic parameters (in particular the intrinsic $\beta-t_c$ degeneracy), while reconstruction of additional polarization modes is governed by the independence and conditioning of the antenna responses.

The marginalized uncertainty on $\beta$, follows from the Schur complement. Defining:
\begin{equation}
r_\beta=\frac{\Gamma_{\beta\lambda}\Gamma_{\lambda\lambda}^{-1}
\Gamma_{\lambda\beta}}{\Gamma_{\beta\beta}}\,,
\end{equation}
from \cref{sigmabetanetwork} one obtains the compact expression:
\begin{equation}
\sigma_\beta=\frac{1}{\sqrt{\Gamma_{\beta\beta}(1-r_\beta)}}\,.
\end{equation}
where $\Gamma_{\beta\beta}$ encodes the intrinsic information on the dispersive phase and $r_{\beta}$ quantifies the fractional information loss due to nuisance-parameter correlations. Here $\Gamma_{\beta\beta}$ depends primarily on the waveform model and detector sensitivity (and is only weakly geometry-dependent at fixed signal to noise ratio (SNR)), whereas the network configuration affects $\sigma_{\beta}$ through the nuisance block $\Gamma_{\lambda\lambda}$ that appears in the Schur complement.

$\sigma_\beta$ can be expressed as in \cref{sigmabetanetwork}:
\begin{equation}
\label{sigmabnetwork}
\sigma_\beta=\left[\Gamma_{\beta\beta}-\Gamma_{\beta\lambda}
\Gamma_{\lambda\lambda}^{-1}\Gamma_{\lambda\beta}\right]^{-1/2}\,.
\end{equation}
The geometrical contribution to $\Gamma_{\lambda\lambda}$ is set by the detector time delays. From \cref{Gammasky} one finds asymptotically $\Gamma^{\rm sky}_{ab}\propto L^2$, hence the corresponding covariance block scales as $C^{\rm sky}_{ab}\propto L^{-2}$. Increasing the baseline therefore raises the eigenvalues of the localization Fisher block and suppresses the associated covariance eigenvalues, which improves the conditioning of the nuisance-parameter block entering the Schur complement. This improvement does not guarantee a decrease of every single correlation coefficient; rather, it reduces the contribution of poorly constrained geometrical combinations to the Schur term, thereby lessening the impact of sky-position uncertainty on the $\beta$ uncertainty. 

A quantitative comparison between different detector layouts requires the inversion of the complete Fisher matrix. The relative performance between the two layouts can be computed introducing the ratio of the uncertainties:
\begin{equation}
\label{rbeta}
R_\beta=\frac{\sigma_\beta^{\rm (2L)}}{\sigma_\beta^{\triangle}}=\left[
\frac{\Gamma_{\beta\beta}^{\triangle}(1-r_\beta^{\triangle})}
{\Gamma_{\beta\beta}^{\rm (2L)}(1-r_\beta^{\rm 2L})}\right]^{1/2}\,.
\end{equation}
Thus the advantage of a double-$L$ network cannot be deduced from baseline scaling alone but must be assessed using the full Fisher inversion (waveform model, sensitivity curves, antenna responses and all parameter correlations). Still, the substantial increase of localization Fisher eigenvalues for a well-optimized long-baseline network typically yields a much better conditioned nuisance block and hence improved marginalized constraints on $\beta$.

We now apply the same Fisher framework to the reconstruction of the scalar (breathing) polarization $h_b$ predicted in $f(R)$ theories. In this case the limiting factor is not a universal baseline scaling but the rank and the conditioning of the antenna-pattern matrix $\mathbf{F}$.
The detector response is:
\begin{equation}
h_k=F_k^+h_++F_k^\times h_\times+F_k^bh_b\,,
\end{equation}
After marginalization over the tensor modes, the uncertainty on the scalar amplitude is:
\begin{equation}
\sigma_{b}=\sqrt{(\Gamma^{-1})_{bb}}\,,
\end{equation}
and the relative performance between detector architectures is:
\begin{equation}
\label{rb}
R_b=
\frac{\sigma_{b}^{\rm (2L)}}{\sigma_{b}^{\triangle}}=\left[
\frac{\Gamma_{bb}^{\triangle}(1-r_b^{\triangle})}
{\Gamma_{bb}^{\rm (2L)}(1-r_b^{\rm 2L})}\right]^{1/2}\,.
\end{equation}
Contrary to dispersive propagation effects, the sensitivity to additional
polarization modes is not governed by a universal scaling with the detector baseline. Instead, it depends on the rank and conditioning of the antenna-pattern matrix:
\begin{equation}
{\mathbf F}=
\begin{pmatrix}
F_1^+ & F_1^\times & F_1^b\\
F_2^+ & F_2^\times & F_2^b\\
\vdots & \vdots & \vdots
\end{pmatrix}\,.
\end{equation}
For the ideal triangular ET the detector tensors satisfy the null-stream identity \cref{nullstream}, hence the instantaneous amplitude-only antenna matrix has algebraic rank$\leq2$. Consequently a single-event, amplitude-only measurement cannot uniquely determine three independent polarization amplitudes. However, waveform evolution and multi-detector timing (and multiband data) can partially lift these degeneracies in a full parameter-estimation analysis. For instance, the geographical separation of the two sites changes the relative orientation of the detector tensors due to the curvature of the Earth. This rotation modifies the antenna-pattern matrix, as illustrated before: ${\mathbf F}_{kA}=D_k^{ij}e^A_{ij}$ and consequently changes the eigenvalues of ${\mathbf F}^{T}{\mathbf F}$. Therefore, the polarization performance of the double-$L$ configuration depends on both the relative detector orientation and the source sky position \cite{Branchesi}. The triangular configuration may provide an advantage because of its symmetry and redundancy, whereas a suitably oriented double-$L$ network can provide comparable or, for specific source locations, improved conditioning of the accessible polarization subspace.

\begin{figure}[htbp]
    \centering
    \includegraphics[width=1\textwidth]{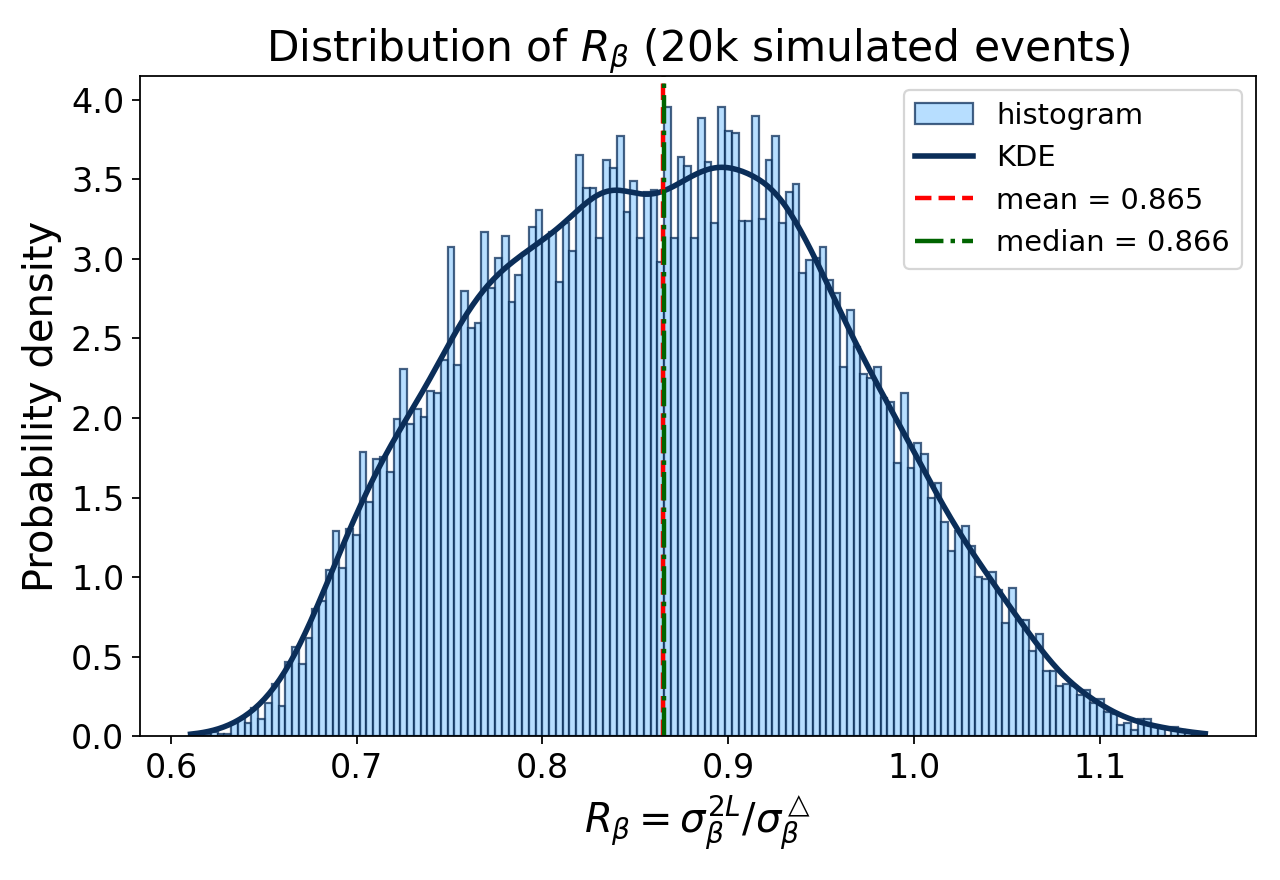}
    \caption{Histogram of the distribution of the $R_{\beta}$ values obtained from 20,000 simulated events with isotropically distributed sky positions, with a mean value of $R_{\beta}=0.87$.}
    \label{fig1}
\end{figure}

The final comparison between the two detector architectures is therefore given by the two ratios $R_{\beta}$ and $R_b$ \cref{rbeta,rb}. By simulating these two quantities, we obtained the metric used to compare the performance of the different configurations. The simulations for $R_{\beta}$ and $R_b$ were carried out by generating a Monte Carlo ensemble of $20{,}000$ events, for each event, the network SNR and the geometric antenna response of each detector were computed separately. 
The network SNR was obtained from a frequency-domain integral weighted by the inspiral amplitude ($\propto f^{-7/6}$), with the overall signal amplitude determined by the chirp mass and luminosity distance, which were drawn from prescribed log-normal distributions. The antenna response was obtained by by contracting the polarization tensors $(e_+,\,e_\times,\,e_b)$ with the Michelson detector tensor for both the triangular configuration  and the double$L$ configuration. From these ingredients two distinct estimators were built: $\sigma_\beta$, obtained by marginalizing over the relevant nuisance parameters using the Schur complement of the full Fisher matrix, and $\sigma_b$, from a full $2\times2$ Fisher matrix on the $(F_+, F_b)$ subspace, numerically inverted with explicit rank and determinant checks to identify geometries for which the network Fisher matrix becomes singular. The nuisance parameters were marginalized using a Schur-complement reduction, after which the $2\times2$ Fisher matrix was inverted to obtain the covariance matrix. The ratios $R_\beta = \sigma_\beta^{\rm 2L}/\sigma_\beta^{\triangle}$ and $R_b = \sigma_b^{\rm 2L}/\sigma_b^{\triangle}$ are the resulting random variables of interest, whose distribution over the full ensemble is summarized via histograms and kernel density estimates (KDE) \cref{fig1,fig3}.

\begin{figure}[htbp]
    \centering
    \includegraphics[width=1\textwidth]{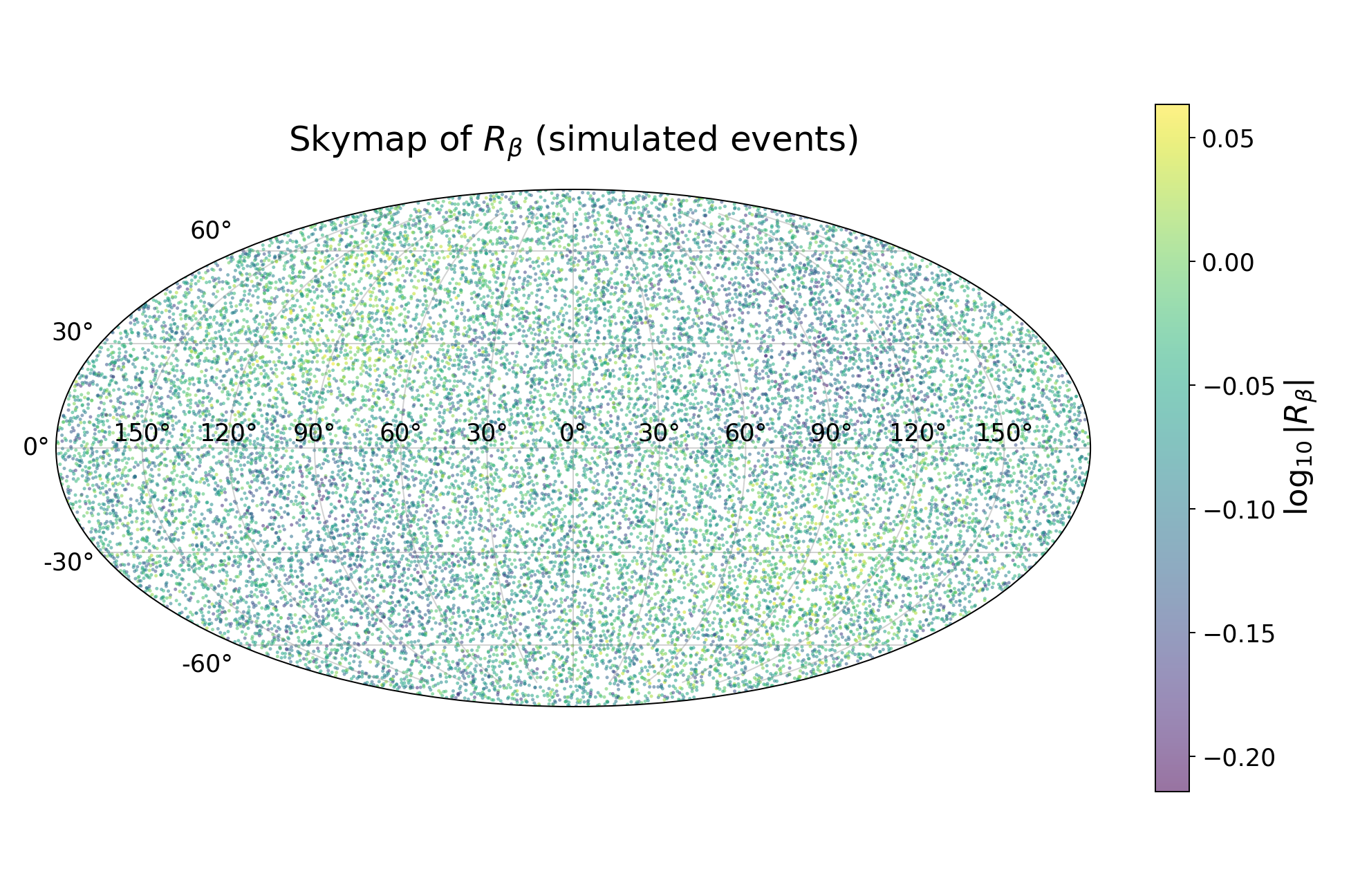}
    \caption{Mollweide sky map of the 20,000 isotropically distributed simulated events, with the color scale indicating the estimated value of $R_{\beta}$ for each event.}
    \label{fig2}
\end{figure}

\begin{figure}[htbp]
    \centering
    \includegraphics[width=1\textwidth]{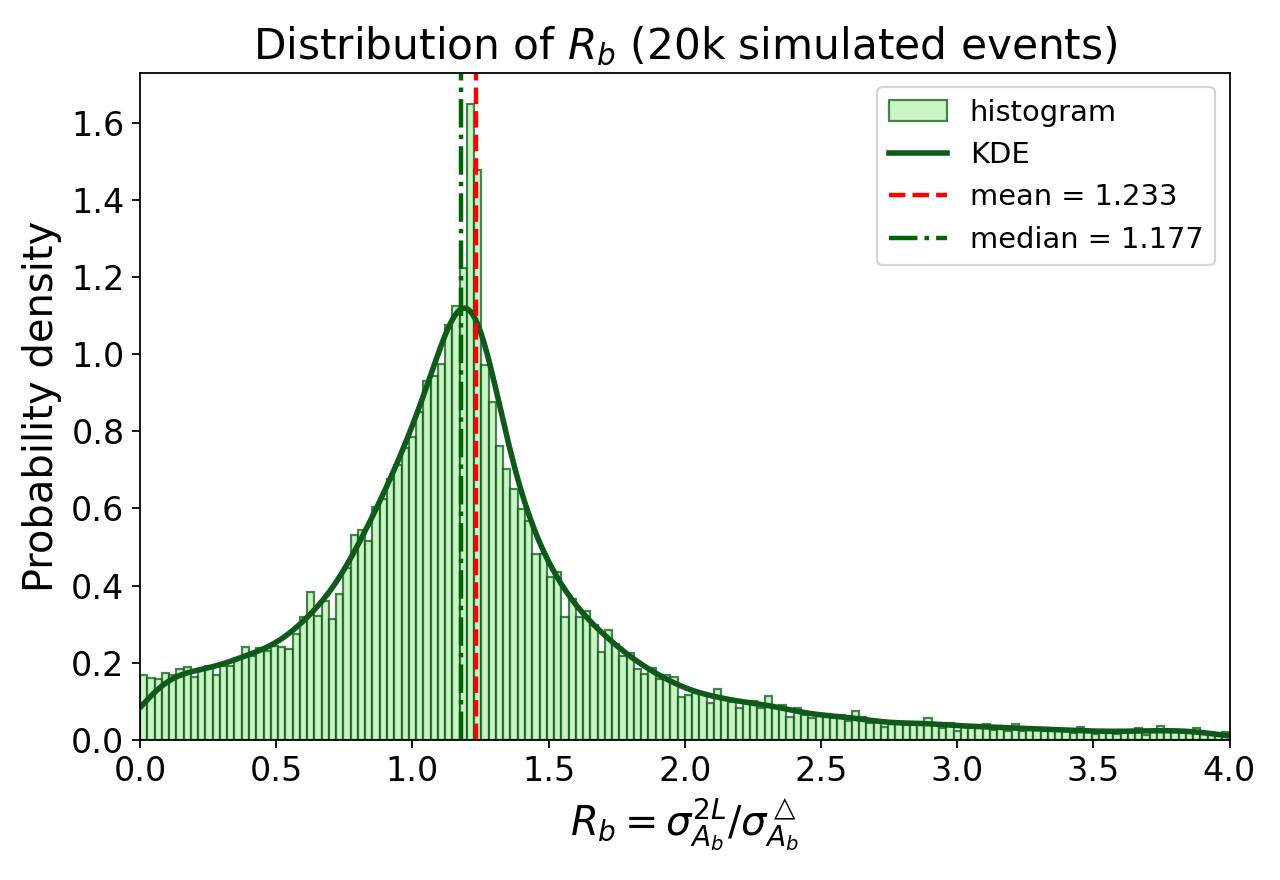}
    \caption{Histogram of the distribution of the $R_{b}$ values obtained from 20,000 simulated events with isotropically distributed sky positions, with a mean value of $R_{b}=1.17$.}
    \label{fig3}
\end{figure}
\begin{figure}[htbp]
    \centering
    \includegraphics[width=1\textwidth]{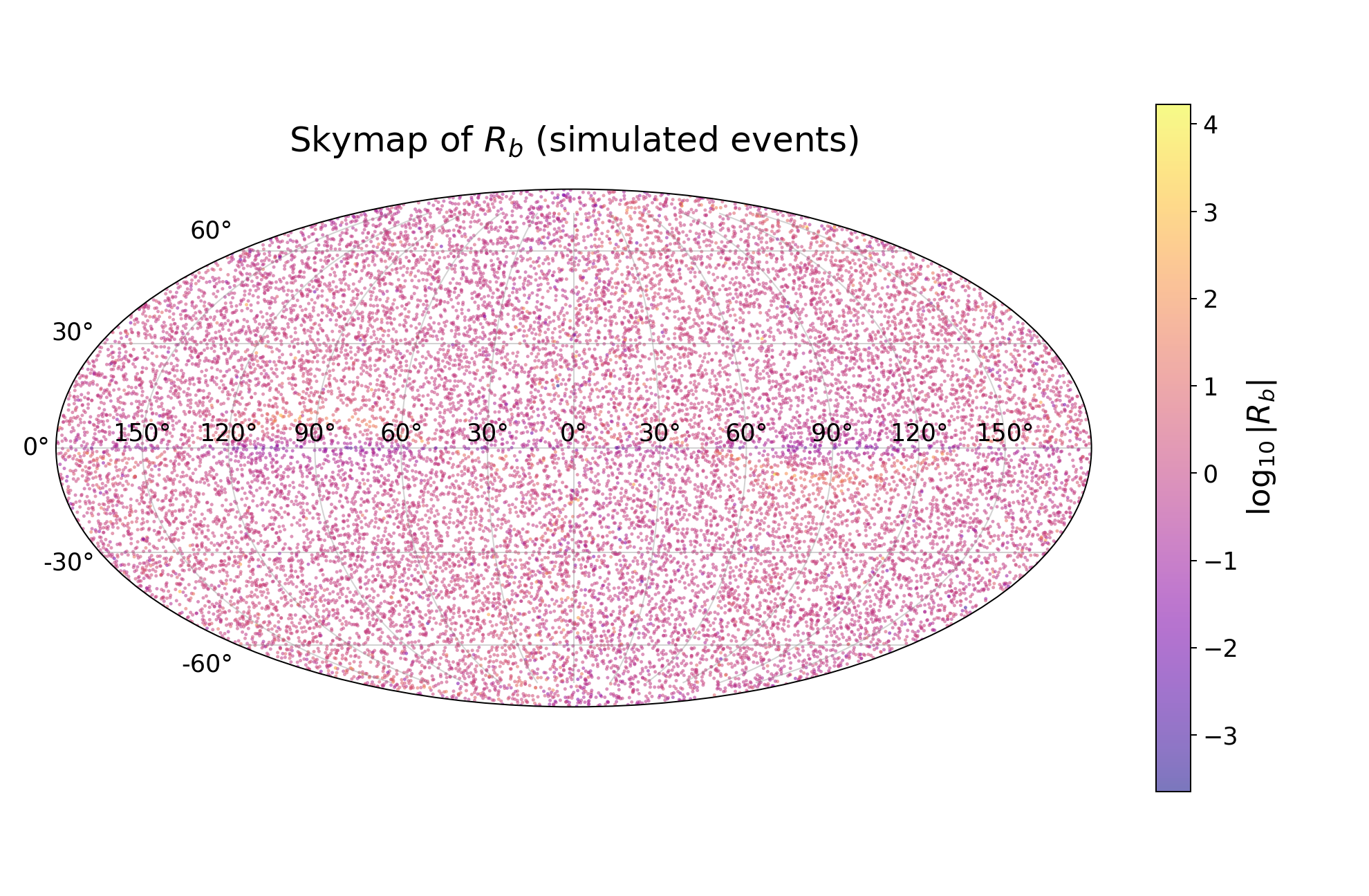}
    \caption{Mollweide sky map of the 20,000 isotropically distributed simulated events, with the color scale indicating the estimated value of $R_{b}$ for each event.}
    \label{fig4}
\end{figure}

The simulated events were distributed isotropically: for each of the $20{,}000$ events, the sky position $(\theta,\phi)$ was sampled isotropically, with $\cos\theta$ uniform in $[-1,\,1]$ and $\phi$ uniform in $[0,\,2\pi]$, while the polarization angle $\psi$ was drawn uniformly in $[0,\,2\pi]$. The resulting distributions of $R_\beta$ \cref{fig1} and $R_b$ \cref{fig3} are therefore not obtained from an analytic average over the antenna pattern, but emerge as ensemble statistics over independently sampled random geometries, allowing the tails associated with unfavorable detector orientations, particularly for the double-$L$ network, to be captured -- tails that a closed-form analytic average would not reproduce. The simulated events used to obtain the distributions of $R_{\beta}$ and $R_{b}$ are shown in \cref{fig2,fig4}, respectively, in Mollweide projection, with a color scale indicating the value of the two simulated quantities.

The final results for the comparison ratios are $R_{\beta}=0.87$ and $R_b=1.17$. These results indicate that the double-$L$ configuration provides, on average, tighter constraints on the dispersive parameter $\beta$ than the triangular configuration, and can therefore also improve constraints on the scalaron mass $m_s$. In particular, the uncertainty $\sigma_{\beta}$ is reduced by approximately $\sim15\%$ with respect to the triangular configuration. Conversely, the triangular configuration proves to be complementary, providing better constraints on the reconstruction of the polarization modes, with an uncertainty $\sigma_b$ that is approximately $\sim15\%$ smaller than that obtained with the double-$L$ configuration. Importantly, this is not a universal event-by-event hierarchy: the two configurations are not always better than one another for every single simulated event, but their relative performance depends on the specific source geometry and detector orientation. The quoted values therefore describe the average behaviour over the full simulated ensemble.

This complementary performance of the two detector configurations is consistent with previous simulation-based studies in the literature \cite{Branchesi,Maggiore}.

\subsection{Implications for viable $f(R)$ models and scalar-mass sensitivity}

The analytical framework developed above can be used to assess which regions of the scalar-mass parameter space associated with viable $f(R)$ models may be accessible to future GW observations. The quantity directly constrained by dispersive propagation is the parameter $\beta=m_s^2D$, rather than the scalaron mass itself. Therefore, for a source at a fixed propagation distance $D$, the sensitivity to the scalar mass is determined by the uncertainty with which the dispersive parameter $\beta$ can be measured. 

The scalaron mass $m_s$ in viable $f(R)$ theories can depend on the background curvature and on the local matter density, particularly in models exhibiting chameleon screening. Consequently, the mass scale relevant for GW propagation may differ from the effective scalar mass in the environment of the source or in the local Universe. The effective scalar masses of some investigated $f(R)$ models span a broad range of values, as already illustrated in \cite{Yang}. The experimental constraints obtained from other observations are of the order of $5.5\times10^{-16}$ eV \cite{Paul}. The projected GW sensitivity considered here should therefore be interpreted as a probe of the parameter region in which the scalaron is sufficiently light to induce an observable dispersive modification of the GW phase, rather than as a direct exclusion limit applicable to all viable $f(R)$ models.

For such a mass, the scalar mode is highly relativistic throughout the frequency band of ground-based detectors. Indeed, the corresponding Compton frequency:
\begin{equation}
f_s=\frac{m_sc^2}{h}\,,
\end{equation}
is approximately
\begin{equation}
f_s\simeq10^{-1}\,{\rm Hz}\qquad(m_s\sim5.5\times10^{-16}\,{\rm eV})\,.
\end{equation}
Although this Compton frequency lies below the main sensitive band of ground-based detectors, the corresponding scalar mode remains highly relativistic and can still produce measurable dispersive phase corrections over long propagation distances. The relevant observable is therefore not the threshold frequency of the scalar mode itself, but the accumulated frequency-dependent phase deformation induced during propagation.

The sensitivity to the scalar mass is consequently determined by the accuracy with which the network can measure $\beta$. The two detector configurations considered in this work affect this sensitivity through their different geometrical information and parameter degeneracies. In particular, a geographically separated double-$L$ configuration can provide stronger timing and geometrical information than a compact triangular configuration, potentially reducing degeneracies between the dispersive parameter and extrinsic parameters such as the source position and propagation delays.

For the same source and under otherwise identical assumptions, the characteristic scalar-mass scales satisfy approximately
\begin{equation}
\label{massratio}
\frac{m_s^{\rm (2L)}}{m_s^{\triangle}}\simeq\left(\frac{\sigma_\beta^{\rm (2L)}}{\sigma_\beta^{\triangle}}\right)^{1/2}=\sqrt{R_{\beta}}\,.
\end{equation}
Thus, an improvement in the marginalized measurement of $\beta$ translates into an extension of the scalar-mass region accessible through dispersive propagation. If the double-$L$ configuration reduces the uncertainty on $\beta$ by a factor:
\begin{equation}
\lambda=\frac{\sigma_\beta^{\triangle}}{\sigma_\beta^{\rm (2L)}}>1\,,
\end{equation}
the corresponding characteristic scalar-mass scale improves by a factor $\sqrt{\lambda}$.

The double-$L$ configuration is therefore expected to be particularly advantageous in the portion of the viable $f(R)$ parameter space where the scalar degree of freedom is sufficiently light to generate a measurable dispersive phase correction, but where the effect is partially degenerate with geometrical delays and other extrinsic properties of the source. The triangular configuration, while providing a different geometrical response and a compact detector topology, remains particularly valuable when the scalar degree of freedom produces an observable additional polarization, since its multiple interferometric data streams and complementary antenna responses can improve the reconstruction of the accessible polarization subspace. The relative performance of the two configurations is therefore expected to depend on the specific observable used to constrain the scalar degree of freedom.

The sensitivity to viable $f(R)$ models is consequently intrinsically multidimensional. It depends on the scalar mass, the source luminosity distance, the corresponding redshift, the signal-to-noise ratio, the chirp mass, the source position and orientation, the detector response, and the correlations among the waveform parameters. In particular, the sky location affects both the antenna-pattern functions and the geometrical time delays, so that the relative performance of the triangular and double-$L$ configurations is expected to vary across the sky. The present calculation should therefore be regarded as a distance-dependent phenomenological Fisher estimate rather than as a full population-averaged forecast.

\begin{figure}[htbp]
\centering
\includegraphics[width=1\textwidth]{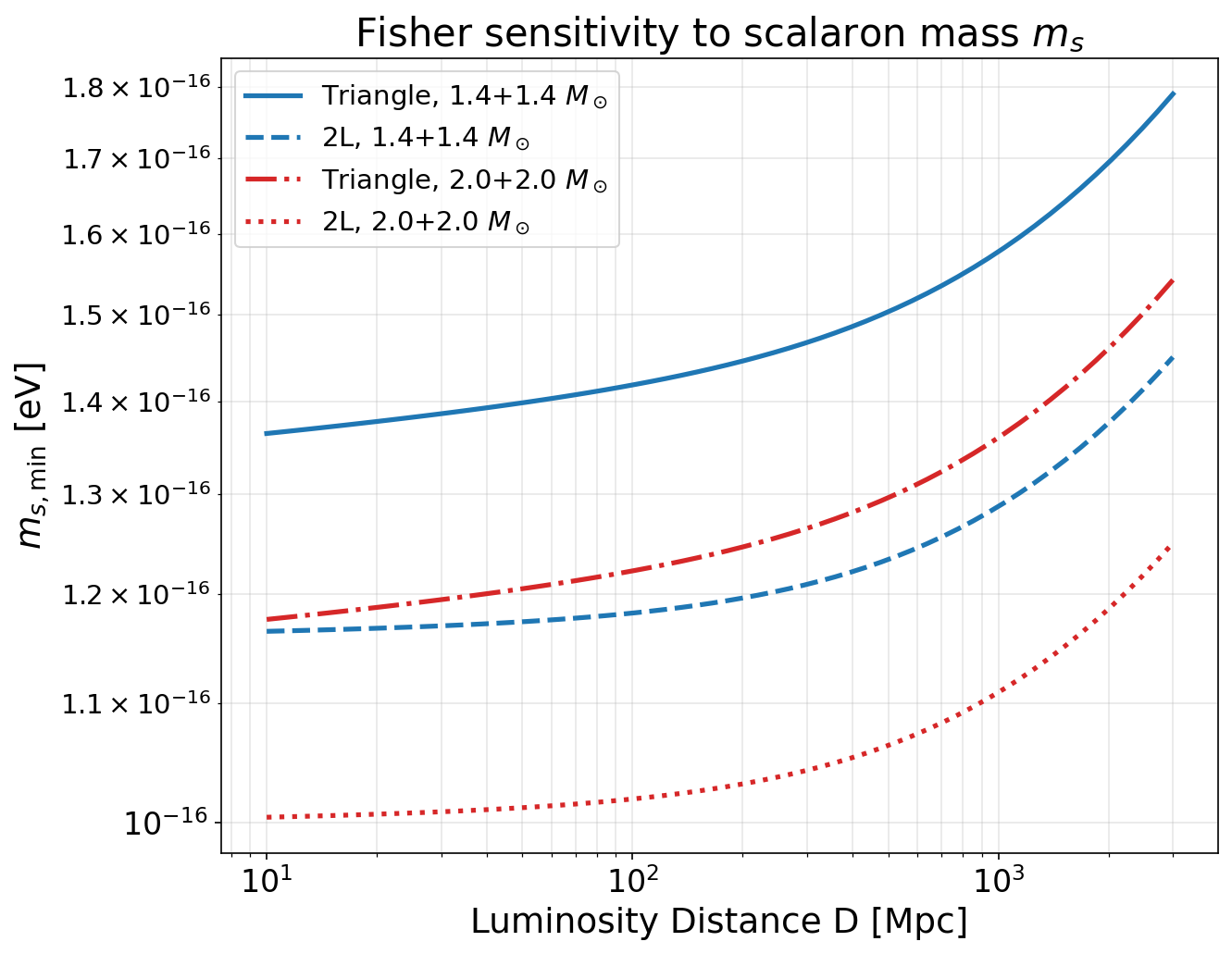}
\caption{Projected scalaron-mass sensitivity for binary neutron-star mergers as a function of luminosity distance and component masses. The sensitivity is obtained from a distance-dependent Fisher analysis including the redshift dependence of the dispersive phase and configuration-dependent parameter correlations. The different geometrical performance is modeled through the effective correlation terms and the performance ratio $R_{\beta}$.}
\label{fig5}
\end{figure}

We estimate the projected sensitivity to the scalar mass $m_s$ by performing a distance-dependent Fisher analysis for compact-binary inspirals. The calculation is performed on a logarithmic luminosity-distance grid $D_L\in[10,3000]\,{\rm Mpc}$. For each luminosity distance, the corresponding source redshift is calculated by numerically inverting the luminosity-distance relation in the adopted flat $\Lambda$CDM cosmology. This redshift is then propagated into the dispersive phase through the factor $(1+z)^{-1}$. In the Fisher analysis adopted here, the dispersive phase is parametrized as:
\begin{equation}
\delta\Psi(f,z)=\beta\,F(f,z)\,,
\end{equation}
where
\begin{equation}
F(f,z)=\frac{1}{4\pi\,(1+z)}f^{-1}\,.
\end{equation}
The factor $(1+z)^{-1}$ accounts for the cosmological redshift dependence of the observed frequency-dependent dispersive correction. Thus, even though the parameterization is written in terms of the distance $D$ entering the definition of $\beta$, the phase response also depends explicitly on the source redshift $z$. In the present calculation, the source redshift is obtained as a function of the luminosity distance, $z=z(D_L)$, by numerically inverting the luminosity-distance relation in a flat $\Lambda$CDM cosmology with: $H_0=70\,{\rm km\,s^{-1}\,Mpc^{-1}}$, $\Omega_m=0.3$, $\Omega_\Lambda=0.7$. The redshift dependence is therefore included directly in the Fisher integrand entering separately through the phase factor $F(f,\,z)$ and is not treated as an independent free parameter in the simplified distance-dependent calculation. This formulation makes explicit that the sensitivity to $\beta$ is controlled simultaneously by the source distance, the chirp mass, the detector noise spectrum, and the redshift-dependent dispersive phase response.

The two representative binary neutron-star systems considered in the calculation are: $1.4+1.4\,M_\odot$ and $2.0+2.0\,M_\odot$. The corresponding chirp masses are computed explicitly from the component masses. The choice of coalescing masses is deliberately restricted to binary neutron stars, since the scalar charge of BHs is expected to vanish or be strongly suppressed in the class of theories considered here by the no hair theorem, whereas neutron stars can carry an effective scalar charge and may therefore provide stronger constraints on $m_s$.

For the frequency-domain inspiral waveform, the Fisher information associated with the dispersive parameter is written as:
\begin{equation}
\Gamma_{\beta\beta}(D,\mathcal{M}_c)=
4\int\frac{|F(f,\,z)|^2|\tilde h(f,\,\mathcal{M}_c,\,D)|^2}{S_n(f)}\,df\,,
\end{equation}
where the Newtonian inspiral amplitude scales as:
\begin{equation}
|\tilde h(f)|\propto\frac{\mathcal{M}_c^{5/6}}{D}f^{-7/6}\,.
\end{equation}
The full numerical calculation evaluates this integral using the frequency-dependent waveform amplitude and the adopted reference power spectral density. In particular, the dispersive phase dependence is included explicitly through $F(f,\,z)$ rather than being represented only by an overall signal-to-noise ratio scaling.

The resulting Fisher element is then converted into the effective marginalized uncertainty. In the present simplified Fisher model, these effects are represented phenomenologically through a configuration-dependent effective correlation factor $r_\beta(D)$ and a geometrical calibration factor $\kappa_{\rm geom}$. The resulting effective uncertainty on $\beta$ is modeled as:
\begin{equation}
\sigma_\beta^{(X)}(D,\mathcal{M}_c)=\frac{\kappa_{\rm geom}^{(X)}}{\sqrt{
\Gamma_{\beta\beta}^{(X)}\left[1-r_\beta^{(X)}(D)\right]}}\,,\qquad
X=\triangle,{\rm 2L}\,,
\end{equation}
where the geometrical calibration factors are chosen considering the averaged parameter $R_{\beta}$ obtained in the previous section. This expression explicitly separates the information supplied by the frequency-domain waveform and the detector noise from the effective degradation or improvement associated with parameter correlations and network geometry. This treatment is motivated by the fact that the precision with which $\beta$ is measured is not determined solely by the instrumental noise level. A network with a more favorable geometry can constrain the extrinsic properties of the source more efficiently, thereby reducing the degeneracy between a dispersive propagation effect and other parameters entering the waveform.

Away from the boundary $m_s=0$, standard error propagation gives:
\begin{equation}
\sigma_{m_s}\simeq\frac{\sigma_\beta}{2m_sD}\,.
\end{equation}
For the projected sensitivity curves discussed below, we impose a nominal threshold on the measurable dispersive parameter and infer the corresponding characteristic scalar-mass scale:
\begin{equation}
\beta_{\min}=k\sigma_\beta\,,\qquad k=3\,,
\end{equation}
which gives
\begin{equation}
\label{msmin}
m_s^{\min}(D)=\sqrt{\frac{k\,\sigma_\beta}{D}}\,.
\end{equation}
The resulting quantity should therefore be interpreted as a characteristic sensitivity scale for the smallest dispersive scalar mass that could produce a phase correction at the adopted significance threshold, rather than as a universal upper or lower bound on the scalaron mass.

An important consequence of this combined treatment is that the sensitivity to $m_s$ degrades monotonically with increasing source distance. Although a longer propagation baseline would, in principle, enhance the accumulated dispersive phase, this effect is exactly compensated by the corresponding loss of signal-to-noise ratio, since the GW amplitude decreases as $D_L^{-1}$. In the present Fisher model,
the uncertainty on $\beta$ scales as $\sigma_\beta\propto D_L(1+z)$, so that the two effects cancel in the ratio entering $m_s^{\rm min}$ (\cref{msmin}). The residual degradation of the projected scalar-mass sensitivity with distance is therefore driven entirely by the cosmological redshift factor $(1+z)^{-1}$ entering the dispersive phase response.

The resulting curves in \cref{fig5} indicate that future ground-based GW observations may improve the constraints on $m_s$ with respect to existing observational limits \cite{Jana,Paul}, potentially by a substantial factor depending on the source population and detector configuration. The precise improvement should not, however, be interpreted as a universal property of the detector topology. It depends on the assumed reference PSD, the source distance and redshift, the binary masses, the sky location and orientation, and the treatment of parameter correlations. Within the simplified phenomenological model adopted here, the double-$L$ configuration provides a systematically improved marginalized measurement of $\beta$ relative to the triangular configuration. Because the scalar-mass scale depends on the square root of the uncertainty on $\beta$, the corresponding improvement in the accessible $m_s$ scale is less pronounced than the improvement in $\sigma_\beta$ itself. The numerical ratio between the two configurations should therefore be interpreted as a forecast of their relative sensitivity under the adopted phenomenological assumptions. From \cref{massratio} and the results obtained in the previous section it is possible to compute the ratio $m_s^{\rm (2L)}/m_s^\triangle=\sqrt{R_\beta}\simeq0.9$, finding an averaged systematical improvement of nearly $\sim8-10\%$ on the scalaron mass constraint obtained by the double-$L$ configuration compared with the triangle layout, as visible in \cref{fig5}.

The definitive comparison would require a full network Fisher parameter-estimation analysis incorporating the complete detector geometry, antenna responses, sky localization, polarization content, distance uncertainty, and a cosmologically consistent treatment of the dispersive propagation kernel.

\section{Conclusion}
\label{conclusion}

Since viable $f(R)$ theories still admit regions of parameter space that are not excluded by current observations, gravitational-wave observations with third-generation detectors provide a unique opportunity to probe additional gravitational degrees of freedom. In this work, we investigated the observational signatures associated with the massive scalar mode predicted by viable $f(R)$ models, considering both propagation and generation effects, including dispersive phase modifications and scalar-tensor arrival-time differences. We developed a unified waveform framework incorporating these effects through the dispersive parameter $\beta=m_s^2D$, providing a direct connection between GW observations and the scalaron mass $m_s$.

We developed an analytical Fisher-information framework to investigate how detector geometry affects the estimation of modified-gravity parameters. By expressing the marginalized uncertainty on $\beta$ through the Schur complement, we explicitly separated the intrinsic dispersive information from the loss of information associated with correlations with extrinsic parameters. We showed that the advantage of geographically separated detector networks originates from the enhancement of geometrical time-delay information and the improved conditioning of the extrinsic-parameter sector. In particular, the localization Fisher information scales approximately as $\Gamma^{\rm sky}\propto L^2$, although the resulting improvement in the marginalized constraint on $\beta$ is more moderate because it depends on the complete covariance structure of the network.

Our simulations quantify the complementary performance of the two detector architectures considered in this work. Our results show that the relative performance of the two detector configurations is event-dependent: neither layout is uniformly superior, as their relative performance depends on the specific sky position and source orientation of each event, with one configuration outperforming the other for some events and underperforming for others. Consequently, a meaningful comparison between the two architectures can only be drawn from their ensemble-averaged performance across the full population of simulated events. The geographically separated double-$L$ configuration provides constraints on the dispersive parameter $\beta$ that are on average approximately $\sim15\%$ tighter than those obtained with the triangular configuration. This improvement is a consequence of the enhanced timing and sky-localization information provided by the $\sim10^3$ km baselines, which help disentangle dispersive propagation effects from correlations with extrinsic parameters. As a consequence, in posing constraints on the scalaron mass $m_s$ the double-$L$ layout outperforms the triangle configuration of nearly $8-10\%$. Conversely, the triangular configuration performs better in the reconstruction of additional polarization modes, providing constraints that are on average approximately $\sim15\%$ tighter than those obtained with the double-$L$ network. This complementary behavior reflects the different physical mechanisms governing the two observables: propagation effects are primarily sensitive to geometrical timing information, whereas polarization reconstruction depends on the rank and conditioning of the antenna-pattern response.

The accessible mass range depends on the source distance, detector sensitivity, waveform model, and frequency range over which the dispersive correction is measurable. Our results indicate that a future third-generation detector such as the ET can improve the experimental constraints on the scalaron mass $m_s$ depending on the configuration considered. If implemented in a geographically separated double-$L$
configuration, the detector can obtain the best experimental results for dispersive-propagation constraints on $m_s$.

Overall, our results show that the optimal detector configuration depends on the target modified-gravity observable. Geographically separated double-$L$ networks are particularly advantageous for propagation-based tests of scalar degrees of freedom and for constraining the scalaron mass, while triangular configurations remain competitive, and can be superior, for the reconstruction of additional polarization modes. This complementarity is especially relevant because different regions of the modified-gravity parameter space may leave their dominant signatures through accumulated dispersive propagation, scalar-tensor arrival-time effects, or additional polarization content. In particular, very light scalar fields can produce measurable dispersive effects accumulated over cosmological distances, whereas scenarios with appreciable scalar radiation or breathing polarization may be better constrained through the polarization response of the detector network.

Finally, multiband observations combining third-generation ground-based detectors with space-based observatories such as LISA could provide complementary information on the binary evolution over a broader frequency range, potentially reducing correlations between astrophysical and modified-gravity parameters and further improving the constraints on additional gravitational degrees of freedom. More generally, our analytical framework provides a physically motivated basis for interpreting the role of detector geometry in tests of gravity beyond General Relativity and for optimizing future GW detector networks according to the specific signatures of modified gravity.

\section{Aknowledgments}
The author would like to acknowledge networking support by the COST Action 23130.







\end{document}